\documentstyle[preprint,aps,epsf]{revtex}
\begin{document}
\draft
\newcommand{\abst}[1]{\hspace{10mm}\begin{minipage}[t]{155mm}#1\end{minipage}\vspace{5mm}}
\newcommand{\T}{\rm{T}}
\newcommand{\sigm}{\Delta x}
%
%
\def\etal{{\it et al~}}
%
%

%
\draft
%
%
%
%
\title{'Superluminal' tunnelling, quantum measurements and the speed of information transfer}
%
%
\author{D.Sokolovski, A.Z.Msezane$^{a)}$ and V.R.Shaginyan$^{b)}$}
\address{School of Mathematics and Physics,
         Queen's University of Belfast, 
	 Belfast, BT7 1NN, United Kingdom\\
$^{a)}$ Department of Physics and Center for Theoretical Studies
of Physical Systems, Clark Atlanta University, Atlanta, Ga., 30314,
USA\\
$^{b)}$ Petersburg Nuclear Physics Institute, Gatchina, 188300,
Russia}
\date{\today}
\maketitle
%
%
\begin{abstract}
We exploit the analogy between the transfer of a pulse  
 across a scattering medium 
 and Aharonov's weak measurements
to resolve the long standing paradox between the 
impossibility
to exceed the speed of light
and the seemingly 'superluminal' behaviour  
of a tunnelling particle in the barrier or a photon in a
 'fast-light' medium. 
We demonstrate that 'superluminality' 
occurs when the value of the duration $\tau$ spent in the
barrier is uncertain,
whereas when $\tau$ is known accurately, no 
'superluminal' behavior is observed.
In all cases only subluminal durations
contribute to the transmission 
which precludes faster-than-light information transfer,
as observed in a recent experiment.

\end{abstract}
%
%
%

Recent experiments \cite{Natur} on transmitting information-containing 
features of an optical pulse across the 'fast-light'
medium, in which the group velocity exceeds the vacuum 
speed of light $c$, have renewed the interest in the so-called
'superluminal' propagation phenomenon.
It is well known that a wavepacket 
transmitted across the potential barrier,
undersized wave guide or fast-light medium 
may arrive in a detector ahead of the one that propagates freely,
as if it has crossed the scatter infinitely fast.
This phenomenon, often referred to as 'apparent superluminality
of quantum tunnelling' was first noticed more than
seventy years ago \cite{McColl}
and often discussed since (for Reviews see \cite{Rev1,Rev2},
a recent selection of different views on the subject can be
found in \cite{Ann}).
It seems to raise the question about the possibility
of faster-than-light travel, e.g., in classically forbidden regions,
inaccessible to a classical particle.
It is, however, broadly understood that
 the paradox results from an 
incorrect identification of the transmitted 
peak with the incident one, since the incident pulse 
undergoes severe reshaping in the barrier region.
Although it has been suggested \cite{Low1}, that a barrier
may, in some manner, carve the transmitted pulse from the forward
tail of the incident one, the question of 
how exactly this is achieved remains open. 
 The precise mechanism of this 
reshaping, its relation to Aharonov's 'weak' measurements 
\cite{Ah1,Ah2,Ah3} of the
time delay 
and the implications for the information transfer,
studied in \cite{Natur}, 
are the subjects of this Letter.  
Earlier work on the weak nature of the tunnelling times  
can be found in \cite{St1}-\cite{Iann},  
and a  more recent approach relating 'superluminality' to
superoscillations is given in \cite{AhS}.   
\newline \indent
We start by revisiting \cite{Ah1,Ah2} the analysis 
of a quantum system 
prepared at $t=0$ in the initial state $|I>$ and then post-selected
 (observed)
at $t=\T$ in the final state $|F>,$
\begin{equation}
|I>=\sum_{\nu} a_{\nu} |{\nu}>\quad |F>=\sum_{\nu} b_{\nu} |{\nu}>,
\quad \hat{A}|{\nu}>=A_{\nu}|{\nu}>. 
\end{equation}
If at some $t$, $0<t<\T$,  
the system is subjected to a von Neumann-type
measurement \cite{Neum} of an operator $\hat{A}$,
the state of the pointer with position $\tau$ after post-selection is given by
\cite{Ah1,Ah2}
\begin{equation} \label{eq:AHAR}
<\tau|M> = \sum_{\nu}G(\tau-A_{\nu})\eta_{\nu} \quad\quad \eta_{\nu} \equiv
b_{\nu}^{*}a_{\nu}
\end{equation}
where $G(\tau)$ is the initial (e.g., Gaussian) state of the meter at $t=0$ and
$\sum_{\nu}$ must be replaced by an integral $\int d\nu$ if
$\hat{A}$ has a continuous spectrum. 
The meter is then read, i.e., the pointer position is accurately
determined.
For an accurate 'strong' measurement,
 the width $\Delta \tau$ is small (compared to the 
separation between the eigenvalues or, if the spectrum is continuous, to the scale
on which $\eta_{\nu}$ varies considerably)
and the meter's readings occur close to the eigenvalues $A_{\nu}$.
For an inaccurate, or 'weak', measurement, $\Delta \tau$ is large,
and the interference between overlapping Gaussians in Eq.(\ref{eq:AHAR})
may, for a special choice of $|I>$ and $|F>$,
produce anomalous readings in the regions where 
 $\hat{A}$ has no eigenvalues \cite{Ah1,Ah2}.
Next we show that when using the coordinate of the tunnelled particle
to estimate the time delay it has experienced in the barrier, $\tau$,
one, in fact, performs a weak measurement of $\tau$.  
The anomalously small value of $\tau$ that is obtained, 
results, just as an Aharonov's weak value, from the  
quantum uncertainty inherent to the procedure.
\newline \indent
Consider a one-dimensional wave packet transmitted across a short-ranged potential 
barrier $V(x)$, contained inside the region $0<x<b$. 
At $t=0$
 the particle is prepared as an incident 
wave packet with a mean momentum $k_0$, centred at 
$x=x_I<0$
\begin{equation} \label{eq:init}
\Psi_0(x)\equiv<x|I>=\int_{-\infty}^{\infty} C(k-k_0)exp(ikx)dk.
\end{equation}
where the factor $C(k-k_0)$ insures that only positive 
momenta contribute to the integral.
At some large $t=\T$, it is post-selected in the transmitted state
($\hbar=1$)
\begin{equation} \label{eq:fin}
\Psi_F(x)\equiv <x|F>=\int_{-\infty}^{\infty} T(k)C(k-k_0)exp[ikx-iE(k)\T]dk,
\end{equation}
where $T(k)$ is the barrier transmission amplitude.
Using the convolution property of the 
Fourier integral, it is convenient to rewrite Eq.(\ref{eq:fin}) as
\begin{equation} \label{eq:fin2}
<x|F>=\int \Psi_{\T}(x-x')\xi(x')dx'
\end{equation}
where 
\begin{equation} \label{eq:env}
{\Psi}_{\T}(x)=\int_{-\infty}^{\infty}C(k-k_0)exp[ikx-iE(k)\T]dk,
\end{equation}
is the state that would evolve from the initial one 
under free propagation, and $\xi(x')$ is the Fourier 
transform of $T(k)$.
We can continue the discussion in terms of the 'tunnelling times'
by identifying ($v_0$ is the velocity corresponding to the wavevector $k_0$)
\begin{equation} \label{eq:class}
\tau(x')\equiv -x'/v_0
\end{equation}
with the delay experienced by the particle in the barrier.
Changing the variables in Eq.(\ref{eq:fin2}) and separating the inessential
phase associated with the free motion,
we obtain
\begin{equation} \label{eq:fin3}
\Psi_F=\exp[ik_0x-iE(k_0)\T] \int G_{}(\tau(x)-\tau)\eta(\tau)d\tau
\end{equation}
where $G$ is the envelope of $\Psi_T$, 
\begin{equation} \label{eq:G}
G(\tau)\equiv \exp[iE(k_0)\T+ik_0 v_0 \tau]
\Psi_{\T}(-v_0\tau)
\end{equation}
and
\begin{equation} \label{eq:eta}
\eta(\tau)\equiv-(2\pi)^{-1}v_0
\int T(k)\exp[i(k_0-k)v_0\tau]dk.
\end{equation}
Comparing Eq.(\ref{eq:fin3}) with Eq.(\ref{eq:AHAR})
shows that the relation between the time delay $\tau$ and
the particle's position $x$ is that between the measured 
quantity, whose amplitude distribution is $\eta(\tau)$
and the position of the pointer, whose initial state is
determined by the envelope of the initial pulse. 
As in the  original Aharonov's approach, the particle is post-selected
in its transmitted state. However, unlike in Ref.\cite{Ah1}-\cite{Ah3},
no external pointer variable is employed and its role
is played by the particle's own position.
Equivalently, registering the transmitted particle at a location $x$ 
amounts to measuring 
the time delay $\tau$ of a particle with the momentum $k_0$.
Finding the particle roughly the the width of the barrier $b$ ahead
of the free one, as it happens in tunnelling \cite{McColl}, corresponds to a negative
time delay of $\approx -b/v_0$.
Importantly,
the accuracy to which the time delayed is evaluated
is limited (if the spreading of the wavepacket is neglected
\cite{FOOT0}) 
 by the uncertainty $\Delta x$ of the particle's position in
its initial state \cite{FOOT1}, $\Delta \tau \approx \Delta x/v_0$.
\newline
In the case of tunnelling, the momentum spread of the initial
wavepacket must be at least \cite{FOOT2} small enough for all its components to tunnel,
rather than to pass over the barrier. Such wavepackets, broad 
 in the coordinate space, correspond to inaccurate weak 
 measurements which may produce anomalous 'superluminal'
 readings, even when the 
 amplitude distribution $\eta(\tau)$ contains only non-negative time delays. 
Indeed, in the complex $k$-plane, $T(k)$ may only have poles
on the positive imaginary axis and in the lower half-plane \cite{Landau}.
The poles of the first kind (I)
correspond to the 
bound states supported by $V(x)$, while those of the second kind (II)  
correspond to scattering resonances.
Closing the contour of integration in Eq.(\ref{eq:eta}) as appropriate,
 we obtain
\begin{eqnarray}\label{eq:init11}
\eta(\tau)=2\pi i\sum_{I}Res_{n}T\exp(ik_nv_0\tau), \quad \tau<0\\
\quad\quad\quad\quad = -2\pi i\sum_{II}Res_{n}T\exp(-ik_nv_0\tau), \quad \tau >0.
\end{eqnarray}
where $Res_{n}T$ denotes the residue of $T(k)$ at the $n$-th pole.
Thus, the poles (I), if present, produce negative time delays and are
 responsible, in the classical limit, for the speed up of a particle 
passing above a potential well.
It is interesting to note
that a potential well too shallow to support a bound state
would not speed up a passing wavepacket.
If, on the other hand, 
no bound states are present, then 
\begin{equation}\label{eq:caus}
\eta(\tau)\equiv 0$ for $\tau<0
\end{equation}
and 
the 'spectrum' of the time delays in Eq.(\ref{eq:fin3}) is confined in the
$0\ge\tau<\infty$ semi-axis.
Note that Eq.(\ref{eq:caus}) demonstrates 
the causal nature of the scattering process, since
the condition $Im k_n <0$, used in its derivation, also ensures
that $ImE(k_n)<0$ for $Rek_n>0$ and the resonance states
containing outgoing waves, $Rek_n > 0$,
 are emptied,
 rather than filled up, as the time increases \cite{Landau}.
 \newline \indent
It is clear now that with a careful choice of $T(k)$,
and, therefore, $\eta(\tau)$
the 'superluminal' pulse can be produced, in an explicitly
causal manner, from the front tails of $\Psi_T(x-x')$, all delayed
relative to free propagation.
One such system is a particle tunnelling across a potential barrier.
However, our approach only relies on the analyticity of 
the transmission amplitude $T(k)$ and can also be applied to 
evanescent propagation in waveguides \cite{Ann}
and optical propagation through 'fast-light' media
(see \cite{Rev1},\cite{Natur}),
which have additional advantage of dealing with wavepackets
not subject to spreading in vacuum.
\newline 
Information transfer is often associated with propagation 
of non-analytic features, such as cut-offs \cite{Natur} and next 
we will show that Eq.(\ref{eq:caus}) ensures that it cannot be
transferred faster than light.
For a simple example, consider the propagation of an electromagnetic pulse
across two narrow semi-transparent mirrors, broadly similar
to the setup studied in \cite{China}. If the mirrors are 
modelled by $\delta$-functions of magnitude $\Omega$ located
at $x=0$ and $x=b$, respectively, $T(k)$ is given by the multiple
scattering expansion
\begin{equation} \label{eq:mult}
T(k)=\sum_{m=0}^{\infty}T^{(m)}(k)\equiv(1+R(k))\sum_{m=0}^{\infty}R(k)^{2m}\exp(2imkb)
\end{equation}
where 
\begin{equation} \label{eq:ind}
R(k)=-i\Omega/(2k+i\Omega)
\end{equation}
is the reflection amplitude for a single $\delta$-function placed at the 
origin $x=0$. Accordingly, the distribution ${\eta(\tau)}$
is decomposed into
subamplitudes $\eta_m(\tau)$, $\eta_m(\tau)\equiv
0$ for $\tau<2mb/c$,
each peaked near $\tau_m = 2mb/c$ (Fig.1a).
For a large $\Omega$,$\Omega b \gg 1$, the widths can be neglected and 
the incident pulse is split into a number
of discrete path modes corresponding to $2m$, $m=0,1,2...$
 additional reflections
experienced by the ray between $x=0$ and $x=b$ \cite{China}, 
and Eq.(\ref{eq:fin}) reduces to Eq.(\ref{eq:AHAR}) 
for a variable  with a discrete spectrum $\{ \tau_m\}$,
\begin{equation} \label{eq:AHAR1}
\Psi_F(x) \approx \sum_{m}G(\tau(x)-\tau_m)T^{(n)}(k_0).
\end{equation}
 A numerical evaluation of $\Psi_F(x)$, for $\Omega b=100$, 
 corresponding to  
a weak Gaussian measurement,
$$G(x)=\exp(-x^2/\sigm^2), \quad
\sigm >b,$$ shows (see Fig1.b) how a set of nearly Gaussian shapes, 
each delayed by $\tau_m>0$, interfere to produce 
 a {\it negative} time delay $\approx -b/c$.
Note that this is the best speed up which can be achieved with 
the model Eq.(\ref{eq:mult})
\cite{FOOT}. 
\newline It is now straightforward to show that
this speed up effect
cannot be used to send information faster than light. 
Aharonov and co-workers have already demonstrated \cite{Ah2}
that a weak von Neumann pointer cannot be used for this purpose.
 Rather, they argued, the meter acts
as a filter, extracting, in a non-trivial manner, the signal,
otherwise hidden by a noise. 
The same argument applies to 'superluminal' propagation.   
If the incident wave packet is chosen to be the rear half of a 
Gaussian with a sharp front, $G_-(x)=[1-\theta(x)]\exp(-x^2/\sigm^2)$,
 Eqs.(\ref{eq:fin3}) and (\ref{eq:caus}) show that the transmitted field will vanish 
outside the causal boundary $x_{B}=c\T+x_I$, or for $\tau<0$, as shown in Fig.1c.
Note that in Fig.1c the narrow spikes near $\tau=\tau_m$ result
from large oscillations of $\eta_m(\tau)$ clearly visible in Fig1.a. 
Equally unsuccessful would be an attempt at superluminal transfer
of information encoded in a sharp cut-off at the rear
of the incident pulse, $G_{+}(x)=\theta(x)\exp(-x^2/\sigm^2)$.
Figure 1d shows that by inspecting the field at $x>c\T$, an observer cannot decide whether the whole Gaussian,
or only its front half was incident on the barrier, and must 
await the arrival of the information-carrying part of the signal.
\newline \indent
Recent experiments, in which a detector was to distinguish between 
a cut ($G_{+}$) and an uncut ($G$) signals, 
\cite{Natur} have shown 
that the information detection time for pulses propagating through the
fast-light medium is somewhat longer than that in vacuum, even though the 
group velocity in the medium is in the highly superluminal regime.
As in Fig.1d, the absence of superluminal information transfer
has a simple explanation. Since 
the medium (in this case, the potassium vapor)
does not bind photons, the transmission amplitude
cannot have poles in the upper half of the $k$-plane and
$\xi(x')$ in Eq.(\ref{eq:fin2}) must vanish for $x'>0$.
The most advanced term in Eq.(\ref{eq:fin2}), $\Psi_T(x)$
contains a cutoff at $x=x_B\equiv cT+x_I$,
and the 'superluminal' ($x>x_B$) part of the transmitted field 
in Fig.1d builds up from the
front tails of $\Psi_T(x-x')$, $x'\le 0$,  unaffected by the cut-off,
and for this reason is the same as the advanced uncut field (cf. Fig.2a of
Ref.\cite{Natur}). 
For $x < x_B$ the field has a complicated shape and
the actual delay in detecting the back face of the pulse
is likely to be caused by its deformation while propagating
through the medium.
To advance the cut-off beyond $x_B$, 
and achieve a truly superluminal information transfer,
one requires a one-dimensional system
capable of supporting 
both bound and scattering states of a photon,
in the way a finite-depth potential well supports 
bound states of an electron.
At present, we are not aware of the existence of such systems,
\newline
In summary, the notion of 'superluminality' in wavepacket propagation
is based on relating the final position $x$ of the transmitted particle to the
time $\tau$ it is supposed to have spent in the scatterer.
Quantally, $\tau$ and $x$ are related as the measured quantity and the
pointer position in a measurement, whose accuracy is determined by
the coordinate width of the pulse.
In cases where apparent 'superluminality' is
observed, e.g., in tunnelling or optical propagation
through 'fast-light' media, the measurement
is inevitably weak.
Even if no negative time delays contribute to the
transmission ($\eta(\tau)\equiv 0$ for $\tau <0$),
it may, therefore, produce an anomalous 
reading by constructing a 'superluminal' pulse from
the front tails of the components of the transmitted
pulse, all delayed relative to free propagation.
Just as an improvement in the accuracy 
destroys anomalous weak values \cite{Ah3},
a choice of a narrow incident pulse destroys superluminal propagation 
by making 
higher incident momenta pass over the barrier, or spill
outside the anomalous dispersion region. As a result, a  
'strong' measurement registers only the 'subluminal'
time delays, as illustrated in Fig.1e for the simple 
model described above.
The contradiction between the impossibility of
faster-than-light travel and observing an apparently 
'superluminal' pulse is, therefore, resolved in a typically quantum mechanical
fashion: when 'superluminality' is present, one does not know the 
delay,
and cannot claim that the duration spent in the scatterer is shorter than $b/c$. Conversely,
when
the delay is known, no 'superluminal' transmission is observed.
The absence of negative virtual delays in optical propagation
does, however, limit the speed of information transfer to $c$ and
below, 
as a non-analytic feature (e.g., a cut-off) of the initial pulse
may only travel as far as the most advanced component in Eq.(\ref{eq:fin2}),
i.e., at most by $c\T$. Beyond this point the field will either vanish
if the front part of the pulse was discarded, or remain identical to
the uncut field if its the rear tail has been removed as
the recent experiments by Stenner {\it et al} \cite{Natur} show.

{\bf Acknowledgements:} Two of us (DS and VS) gratefully acknowledge
the support of
the CTSPS, Clark Atlanta University. 
AZM was supported by the U.S. DoE, Division of Chemical Sciences,
Office of Basic Research.

\newpage
\begin{figure}[ht]
\epsfxsize=13cm
\centering\leavevmode\epsfbox{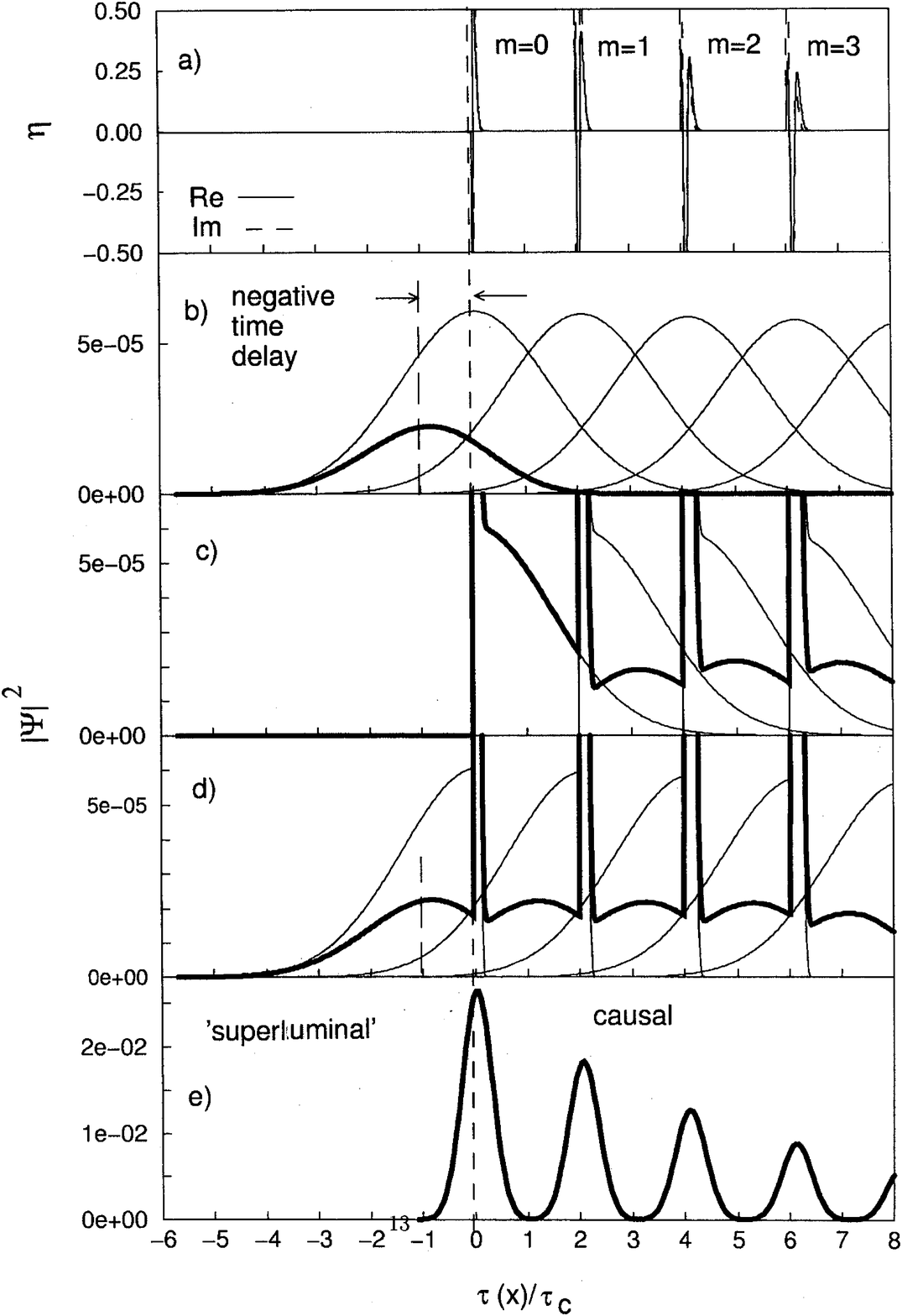}
\vspace{0pt}
\caption{ }
\label{fig:FIGN}
\end{figure}

\newpage
\begin{center}
{\bf Figure captions}
\end{center}

{\bf Fig.1: }
\newline
a)
Real (solid) and imaginary (dashed)
part of the time delay amplitude distribution,
$\eta(\tau)$, for $\Omega b=100$; 
\newline
b) Transmitted field $\Psi_F(x)$ (thick solid) and its components
$\Psi_m$, corresponding to different terms in Eq.(\ref{eq:mult}), for 
$\Omega b=100$, and a Gaussian incident pulse
  $k_0b=1.4\pi$ and $\sigm/b=2.85$;
\newline
c) Same as b)but for an incident Gaussian pulse truncated at front.
\newline   
d) Same as b)but for an incident Gaussian pulse truncated at rear.
\newline
e) Same as b) but for $k_0b=7\pi$ and $\sigm=0.57$.


\begin{thebibliography}{8.}
\bibitem{FOOT0}
For a massive particle, the wavepacket would spread and the
accuracy of the measurement would deteriorate further, as
the time progresses. We, however, assume that the transmitted 
particle is detected before significant spreading has occurred.
No such restriction is necessary for a photon,
whose wavepacket retains its shape in free propagation.
\bibitem{FOOT1}
Consider, for example, free motion, in which case
$\eta(\tau)\approx v_0^{-1}\delta(\tau)$ and 
different values of $\tau(x)$ arise solely from the spread
in the particle's initial position.
\bibitem{FOOT2}
A more stringent condition, $\partial_{k} Re \{ln T(k_0)\}/\Delta x << 1$
is needed if one also wants to avoid distortion.
\bibitem{FOOT}
For $\Omega b\gg 1$ and $k\ll \Omega$, the transmission probability
is small, $|T(k)|\ll 1$, except near equally spaced resonances
at $k_j=\pi j/b$, $j=1,2,..$. Between the resonances we have
($\phi'\equiv \partial_k\phi$) $\phi'(k)\approx -b$, $\phi''(k)\ll 1$.
Thus for the transmitted pulse with $k_0\approx \pi(j+1/2)$ we obtain 
$<x|F>\approx \exp[ik_0x-iE(k_0){\T}]T(k_0)\int[iqx+iq\phi'(k_0)]
\approx T(k_0){G}_T(x+b)$
so that $\tau=-b/c$ is the best speed up achieved with this model.
In addition, we must ensure that the 
momentum width of the incident pulse is small compared
to the spacings between the resonances, $\Delta k \ll \pi$. 
If this is not the case, the wings of the momentum distribution
are caught in the neighbouring resonances, and the barrier
loses its 'filtering' property, transmitting, instead of a single 
pulse, an extended wave, sustained by the decay of the resonances.
The minimal coordinate width of a Gaussian pulse is, therefore,
limited by the condition $4\pi\sigm/b\gg 1$.
\end{thebibliography}
\end{document}